\documentstyle[preprint,aps]{revtex}
\def\umod#1{\ifmmode #1\else $#1$\fi}

\def\beqa{\begin{eqnarray}}
\def\beq{\begin{equation}}

\def\eeqa{\end{eqnarray}}
\def\eeq{\end{equation}}

\def\Eq#1{Eq.(\ref{#1})}

\def\nn{\nonumber}

\def\P{{\cal P}}

\def\sthWsq{\sin^2\theta_{\rm W}}

\def\tthWsq{\tan^2\theta_{\rm W}}
\def\tthW{\tan\theta_{\rm W}}
\def\vex#1{\vbox{\kern#1 pt}}

\def \beq{\begin{equation}}
\def \eeq{\end{equation}}
\def \beqa{\begin{eqnarray}}
\def \eeqa{\end{eqnarray}}
\def \Eq#1{Eq.(\ref{#1})}
\def \J{{\cal J}}

\def \nn{\nonumber}

\def\Cite#1{\thinspace\cite{#1}}
\def\citno#1{\csname b@#1\endcsname}

\input epsf.tex
\begin{document}
\begin{titlepage}
\hfill TPJU 6/96, BNL-
\begin{center}
{\large \bf STRUCTURE FUNCTIONS OF ELECTROWEAK BOSONS AND LEPTONS%
\footnote{Presented by J. Szwed at the Lake Louise Winter 
Institute, Lake Louise, Canada, Feb. 18-24, 1996. Work supported by
the Polish State Committee for Scientific Research
(grant No. 2~P03B~081~09) and  
the Volkswagen Foundation. 
}
}  

\centerline{ \bf Wojciech S\L{}OMI\'NSKI$^b$ and Jerzy SZWED$^{a,b}$%
\footnote{J. William Fulbright Scholar.}}
\centerline{$^a$Physics Department, Brookhaven National Laboratory, Upton, NY 11973, USA}
\centerline{$^b$Institute of Computer Science, Jagellonian University,}
\centerline{Reymonta 4, 30-059 Krak\'ow, Poland}
\end{center}
\begin{abstract}
The QCD structure of the electroweak bosons  is reviewed
and the lepton  structure function is defined and calculated. 
The leading order splitting functions of electron into quarks are extracted,
showing an important contribution from $\gamma$-$Z$ interference. Leading logarithmic
QCD evolution equations are constructed and solved in the asymptotic region where 
 log$^2$ behaviour of the parton densities is observed. Possible applications with 
clear manifestation of 'resolved' photon and weak bosons are discussed.
\end{abstract}
\end{titlepage}  
  
%
%
We would like to present a problem on the edge of the electroweak and strong
sectors of the Standard Model. 
From the point of view of this Institute the question is how far the 
electroweak objects get poluted by the QCD partons. It is known since long
 that 
the QCD structure of the photon ('resolved' photon)\Cite{resph} 
is now clearly visible
in high energy experiments\Cite{resphexp}. Theoretical analysis of the
quark-gluon component 
has been recently extended to weak bosons\Cite{WS1}. The transverse W and Z
develop similar structure with some noticable differences  
 ---  the calculated densities 
show  strong spin and flavour dependence. In the experiment
however one has rarely real weak bosons at one's disposal 
and in most experiments even the high energy photons are  'nearly on-shell' only.
In physical processes, where the structure
of electroweak 
bosons may contribute significantly,  it is the lepton, initiating
the process which is the source of the intermediate bosons. The standard procedure applied in
such cases is to use
the equivalent photon approximation\Cite{WW} (extended also to the case of  weak 
bosons\Cite{WittRoll}) and, 
as a next step, to convolute the obtained boson distributions with parton densities inside
the bosons. For example the parton $k$ density of the electron $F^{e^-}_k$  
would read:
\beq
F^{e^-}_k(z,\hat Q^2,P^2)=\sum_B F^{e^-}_B(\hat Q^2)\otimes F^B_k(P^2)
\label{FF}
\eeq 
where
$(F\otimes G) (z) \equiv \int dx\,dy\,\delta(z-xy) F(y) G(x)$,
$\hat Q^2$ is the maximum allowed virtuality of the boson, 
$P^2$ is the hard process scale and $z$ - the momentum fraction of the parton $k$ with respect to the
electron (detailed definitions follow).
In such an approximation several questions arise:
 how far off-shell can  the intermediate bosons be,
 how large are their interference effects,
 what are the relations between
the energy scales governing the consecutive steps, is the convolution \Eq{FF}
correct in general? 
We think that it is more precise to answer the direct question: 
what is the quark and gluon content
of the incoming lepton or, in other words, what is the lepton structure function? 
The answer to the   above question\Cite{WS22} brings several corrections to the
standard procedure. 

The construction of the lepton structure function can be divided into two
steps. In the first we calculate the splitting of the lepton into a $q\bar q$
pair, in the second the quark-gluon cascade is resummed with the use of the 
evolution equations\Cite{AP}.
To obtain the electron splitting functions 
let us consider inclusive scattering of a virtual gluon off an electron.
In the lowest order in the electromagnetic and strong coupling constants
($\alpha$ and $\alpha_{\rm s}$) the electron 
couples to $q\bar q$ pair as  shown in Figure 1. 
The incoming electron $e$ carries
4-momentum $l$ and the off-shell gluon $G^*$ of 4-momentum $p$ with
large $P^2 \equiv -p^2$, serves here as a probe of the electron.
In the final state we have a massless quark $q$ and antiquark $\bar q$ of 4-momenta $k$ and $k'$
and lepton $\ell'$ (electron or neutrino) of 4-momentum $l'$.
The exchanged boson
$B = \{ \gamma,Z,W \}$ carries  4-momentum $q$ ($Q^2 \equiv -q^2$).
  
The current matrix element squared for an unpolarized electron 
reads:
\begin{eqnarray}
&&\J(e^- \, G^* \rightarrow \ell'\, q_\eta\,\bar q_{-\eta}) =
{\J}_{\mu\nu}^{\eta}(l,p) =\nonumber\\
&&{1 \over 4 \pi}
\int  d\Gamma_{l'} d\Gamma_{k} d\Gamma_{k'} (2\pi)^4 \delta_4(k+k'+l'-p-l) \nonumber\\
&&\times \langle e^-| J_\mu^\dagger(0)|\ell' q_{\eta} \bar q_{-\eta}\rangle
\langle \ell' q_{\eta}\bar q_{-\eta}|J_\nu(0)|e^-\rangle \,,
\label{Jmunueta}
\end{eqnarray}
   where
\beq
 d\Gamma_{k} = {d^4k \over (2\pi)^4} 2\pi \delta(k^2)\,.
\eeq 

For massless quarks we can decompose the current in the helicity basis:

\begin{equation}
{\J}_{\sigma}^{\eta}(l,p) =
\epsilon_{(\sigma)}^{\mu *}(p)\, {\J}_{\mu\nu}^{\eta}(l,p)\, \epsilon_{(\sigma)}^\nu(p)\,,
\end{equation}
where  $\epsilon_{(\sigma)}^\mu(p)$ are polarization vectors of
a spin-1 boson with momentum
$p^\mu = (p_0,0,0,p_z)$:
\begin{eqnarray}
	\epsilon_{\pm}^\mu&=& {1\over\sqrt 2} (0,1,\pm i,0)\,,\\
	\epsilon_0^\mu(q)&=& {\sqrt{1\over |p^2|}}\,(p_z,0,0,p_0)\,.
\end{eqnarray}
In a frame where $\vec q$ is antiparallel to $\vec p$ contributions from different
helicities of exchanged bosons do not mix and the current reads:
\def\T{Q^2}
\beqa
{\J}_{\sigma}^{\eta}(p,l) &&=
 {\alpha_{\rm s} \alpha^2 \over 2\pi}
 \sum_{A,B,\rho }
 g^{Aq}_\eta g^{Bq}_\eta
\int {dy\over y} 
  P^{e^-}_{A_\rho B_\rho}(y)\nn\\
\times&&
\int_{\T_{\rm min}}^{\T_{\rm max}} {\T d\T \over (\T + M_A^2)(\T + M_B^2)}
\,
  H_{\rho\sigma}^{\eta}(x,\T)
\,,
\label{J1}
\eeqa
where $P^{e^-}_{A_\rho B_\rho}(y)$ describes weak boson emission from the electhe $\gamma$-$Z$ interference.
As demonstrated
below their contribution is substantial.

To answer our
main problem of `an electron splitting into a quark' 
we take the limit $\T \ll P^2$ and keep the leading terms only.
Within this approximation the kinematic variables $x,y,z$ read
\beq
y = {pq\over pl},\;\;
z = xy = {-p^2\over 2pl} ,
\eeq
aquiring the parton model interpretation of the quark momentum fraction ($z$)
and of boson momentum fraction ($y$), both with respect to the parent 
electron. 
The leading term of the hadronic part does not depend on the quark helicity~$\eta$:
\beqa
H_{\pm\mp}(x,\T)&=& x^2 \log{P^2 \over \T},\\
H_{\pm\pm}(x,\T)&=& (1-x)^2 \log{P^2 \over \T},
\eeqa
with other components finite for $P^2/\T \rightarrow 0$.

We also recognize $P^{e^-}_{A_\rho B_\rho}(y)$ as a generalization of the splitting functions 
of an electron into bosons\Cite{WS1}:
\beq
P^{e^-}_{A_\rho B_\rho}(y) 
=
 {1\over 2}(
  g^{A{e^-}}_- g^{B{e^-}}_- yY_{-\rho}
+ g^{A{e^-}}_+ g^{B{e^-}}_+ yY_{\rho}
)
\eeq
with
\beqa
Y_+(y) &=& {1 \over y^2}\,,\\
Y_-(y) &=& {{(1-y)}^2 \over y^2}\,,\\
Y_0(y) &=& {2{(1- y)} \over y^2}\,,
\eeqa
where $g^{Ae^-}_\pm$ is the electron  to boson $A_\rho$ coupling in the units of proton charge~$e$.

From kinematics $y \in [z,1-{\cal O}(m_e^2/P^2)]$ and the integration limits for $\T$ read
\beq
\T_{\rm min}= m_e^2 {y^2\over 1-y}\,,
 \;\; 
\T _{\rm max}= P^2{z + y -zy \over z}\,,
\eeq
with $m_e$ being the electron mass. Although smaller than the
already neglected quark masses, it is the electron mass which must be kept
finite in order to regularize collinear divergencies. The upper 
limit of integration 
requires particular attention. In general it is a function of $P^2$, however
integration up to the maximum kinematically allowed value $\T_{\rm max}$
would violate the condition $\T/P^2 \ll 1$. For our approximation
to work we integrate over $\T$ up to $\hat \T_{\rm max}= 
\epsilon P^2$  where $\epsilon \ll 1$ and generally depends on $y$ and $z$.
A similar condition  is in fact  used in phenomenological applications
of the equivalent photon approximation\Cite{Fritj}.  Note that the maximum
virtuality $\hat Q^2_{\rm max}$ can also be kept  independent of $P^2$ when 
special kinematical cuts are arranged in experiment. We concentrate here on 
the fully inclusive  case with the maximum virtuality being $P^2$-dependent.  
Integrating \Eq{J1} over $\T$ within such limits and keeping only 
leading-logarithmic terms
leads to
\beqa
&&{\J}_{\sigma}^{\eta}(p,l)
 =
 {\alpha_{\rm s} \alpha \over 6}
 \sum_{A,B,\rho} 
  g^{Aq}_\eta  g^{Bq}_\eta
 F^{e^-}_{A_\rho B_\rho}(P^2)
\nn\\
&&\otimes
[ P^\rho_{q_\eta } \delta_{\eta,-\sigma} +  P^\rho_{\bar q_{-\eta}} \delta_{\eta,\sigma}]
\log P^2
\,.
\eeqa

In the above equation
$P^\rho_{q_\eta }(x)$ and $P^\rho_{\bar q_{\eta}}(x)$
are boson-quark (-antiquark)
splitting functions\Cite{resph,WS1}
\beq
P^\pm_{q_\pm }(x) =  P^\pm_{\bar q_{\pm}}(x) = 3x^2,\,
P^\mp_{q_\pm }(x) =  P^\mp_{\bar q_{\pm}}(x) = 3 (1-x)^2
\eeq
and $F^{e^-}_{A_\rho B_\rho}(y,P^2)$ is the density matrix of polarized bosons inside electron.
Its transverse components read
\begin{mathletters}
\label{boson}
\beqa
  &&F^{\rm e^-}_{\gamma_\pm \gamma_\pm}(y) =
  {\alpha \over 2\pi}
  {(1-y)^2 + 1 \over 2y}
  \;
  \log \mu_0
\,,\\
  &&F^{\rm e^-}_{{\rm Z}_+ {\rm Z}_+}(y) =
  {\alpha \over 2\pi}
  \tthWsq \; {\rho_{\rm W}^2 (1-y)^2 + 1 \over 2y}
  \;
  \log\mu_Z
\,,\\
  &&F^{\rm e^-}_{{\rm Z}_- {\rm Z}_-}(y) =
  {\alpha \over 2\pi}
  \tthWsq \; {\rho_{\rm W}^2 + (1-y)^2   \over 2y}
  \;
  \log\mu_Z
\,,\\
  &&F^{\rm e^-}_{\gamma_+ {\rm Z}_+}(y) =
  {\alpha \over 2\pi}
  \tthW \; {\rho_{\rm W} (1-y)^2 - 1 \over 2y}
  \;
  \log\mu_Z
\,,\\
  &&F^{\rm e^-}_{\gamma_- {\rm Z}_-}(y) =
  {\alpha \over 2\pi}
  \tthW \; {\rho_{\rm W} - (1-y)^2   \over 2y}
  \;
  \log\mu_Z
\,,\\
  &&F^{\rm e^-}_{{\rm W}_+ {\rm W}_+}(y) =
  {\alpha \over 2\pi}
  {1\over 4\sthWsq} \; {(1-y)^2 \over y}
  \;
  \log\mu_W
\,,\\
  &&F^{\rm e^-}_{{\rm W}_- {\rm W}_-}(y) =
  {\alpha \over 2\pi}
  {1\over 4\sthWsq} \; {1 \over y}
  \;
  \log\mu_W
\,,
\eeqa
\end{mathletters}
where
\beq
  \rho_{\rm W} =  {1\over 2\sthWsq} -1 \,,
\eeq
$\theta_W$ is the Weinberg angle and
\beq
\log\mu_0 = \log {{\epsilon P^2}\over m_e^2}, \;
\log\mu_B = \log{{\epsilon P^2 +M_B^2}\over M_B^2}.
\label{logs}
\eeq
All other density matrix elements (containing at least one  longitudinal
boson) do not develop logarithmic behaviour.

At this point we are able to define the
splitting functions of an electron into  
a quark
at the momentum scale $P^2$ as
\beq
  {\cal P}^{\rm e^-}_{q_\eta}(P^2) =
  \sum_{AB}  g^{Aq}_\eta  g^{Bq}_\eta
  \sum_\rho F^{\rm e^-}_{A_\rho B_\rho} (P^2)\otimes P^\rho_{q_\eta}\,.
\label{sf-def}
\eeq
The expicit expressions for quarks read
\def\rW{\rho_{\rm W}}
\begin{mathletters}
\label{spf}
\beqa
  \P^{\rm e^-}_{q_+}(z,P^2) &&=  {3\alpha \over 4\pi} \{
  e_q^2  \left[ \Phi_+(z) +  \Phi_-(z) \right]  \log\mu_0\nn\\
+ e_q^2 &&\tan^4\theta_{\rm W} \left[ \Phi_+(z) + \rW^2 \Phi_-(z) \right]  \log\mu_{\rm Z}\nn\\
- 2e_q^2 &&\tan^2\theta_{\rm W} \left[ -\Phi_+(z) + \rW \Phi_-(z) \right]  \log\mu_{\rm Z}
\} \,,\\
  \P^{\rm e^-}_{q_-}(z,P^2) &&=  {3\alpha \over 4\pi} \{
  e_q^2  \left[ \Phi_+(z) +  \Phi_-(z) \right]  \log\mu_0\nn\\
+ z_q^2&& \tan^4\theta_{\rm W} \left[ \Phi_-(z) + \rW^2 \Phi_+(z) \right]  \log\mu_{\rm Z}\nn\\
+ 2e_q z_q&& \tan^2\theta_{\rm W} \left[ -\Phi_-(z) + \rW \Phi_+(z) \right]  \log\mu_{\rm Z}\nn\\
+&& (1+\rW)^2 \Phi_+(z) \delta_{q \rm d}  \log\mu_{\rm W}
\} \,,
\eeqa
\end{mathletters}
where
\beqa
\Phi_+(z) &=& {1-z \over 3 z} (2+ 11z +2z^2) + 2(1+z)\log z\,,\\
\Phi_-(z) &=& {2 (1-z)^3 \over 3 z}\,,
\eeqa
and
\beq
z_q = {T_3^q \over \sthWsq} - e_q \,,
\eeq
with $e_q$ and $T_3^q$ being the quark charge and 3-rd weak isospin component, respectively.
The splitting functions for antiquark of opposite helicity can be obtained from 
\Eq{spf} by interchanging $\Phi_+$ with $\Phi_-$.

The splitting functions introduced above show two new features. The first one, 
already mentioned before, is
the contribution from the interference of  electroweak bosons ($\gamma$ and 
$Z$ only). 
The second is their $P^2$ dependence, which arises from the upper integration limit
$\hat \T_{\rm max}$. 

Having completed the first step of our procedure --- the calculation of the 
splitting functions, we can proceed to the resumation of the QCD cascade
using the evolution equations.
 We consider the evolution equations 
in the first order in electroweak couplings and leading logarithmic in QCD. 
Introducing $t=\log(P^2/\Lambda_{\rm QCD}^2)$ we can write (remembering that there is no
direct coupling of the electroweak sector to gluons):
\begin{mathletters}
\label{master}
\beqa
{{dF_{q_{\eta}}^{e^-}(t)}\over dt} &=& {\alpha\over  2\pi} 
\P_{q_\eta}^{e^-}(t)
       + {\alpha_{\rm s}(t)\over 2\pi} \sum_{k,\,\rho} P_{q_\eta}^{ k_\rho} \otimes
F_{k_\rho}^{e^-}(t)
\\
\label{master1}
{{dF_{G_{\lambda}}^{e_-}(t)}\over dt} &=&
 {\alpha_{\rm s}(t)\over 2\pi} \sum_{k,\,\rho} P_{G_\lambda}^{ k_\rho} \otimes
F_{k_\rho}^{e^-}(t).
\label{master2}
\eeqa
\end{mathletters}
We stress that the convolution  of the equivalent boson distributions and boson-quark
splitting functions, \Eq{sf-def}, occurs at the level of splitting 
functions. It is not equivalent to the usually performed convolution of the 
distribution functions because of the $P^2$ dependence of the boson distribution functions \Eq{boson}. Only in the case when the upper limit of integration 
$\hat \T_{\rm max}$ is kept fixed ($P^2$-independent), 
e.g. by special experimental cuts, are the convolutions
equivalent at both levels.


The  equations \Eq{master} can be solved in the asymptotic $t$ region where
we approximate the strong coupling constant by
\begin{equation}
\alpha_{\rm s}(t) \simeq {2 \pi \over bt}\,,
\label{astr}
\end{equation}
with $b=11/2 - n_{\rm f}/3$ for $n_{\rm f}$ flavours.
The asymptotic (large $t$) solution to Eqs.(\ref{master}) for the
parton $k$ of polarization $\rho$ can be now parametrized as
\begin{equation}
F_{k_{\rho}}^{e^-}(z,t) \simeq {1\over 2} 
\left( {\alpha \over 2\pi} \right)^2 
f^{\rm as}_{k_{\rho}}(z) \, t^2
\label{asdef}
\end{equation}
resulting in purely integral equations
\begin{equation}
f^{\rm as}_{i_{\rho}} = \hat\P_{i{_\rho}}^{e^-}
    + {1\over 2b} \sum_{k,\lambda} P_{i_\rho}^{k_{\lambda}}
 \otimes f^{\rm as}_{k_{\lambda}} \,,
\end{equation}
where $\hat\P_{i{_\rho}}^{e^-}(z)$ are given by Eqs.(\ref{spf}) with all $\log\mu_A \equiv 1$.

Numerical solutions  to the above equations with the method described
in Ref.\thinspace[\citno{WS1}] are presented 
in Figure 2 for the unpolarized quark and gluon distributions. 
One notices significant contribution from the  $W$  
intermediate state in the d-type quark density.
The most surprising however is the $\gamma$-$Z$
interference contribution  which cannot be neglected, as it is comparable 
to the
$Z$ term. It violates the standard probabilistic approach
where only diagonal terms are taken into account.  This also stresses 
 the necessity  
of introducing the concept
of electron structure function in which all 
contributions from intermediate bosons are properly summed up.
Due to the nature of weak couplings they 
turn out to be nonzero, 
even in the case of gluon distributions. Again the $\gamma$-$Z$ interference
term  is important and the $W$  contribution dominates in the asymptotic region.  



One should keep in mind that at finite $t$ the logarithms multiplying the
photon contribution differ  from the remaining ones (\Eq{logs}). Being scaled
by $m_e$, they lead to the photon domination at presently available $P^2$. The 
importance of the interference term remains constant relative to the $Z$ contribution,
 as they are both governed by
the same logarithm. But even at presently available momenta, where the 
`resolved' photon dominates, one can see how the correct treatment of the 
scales changes the evolution. In Fig.\thinspace 3 we present the asymptotic solutions
of the evolution equations 
following from our procedure (ESF) compared to those following from naive 
application of the convolution \Eq{FF} (FF). It is possible that 
the difference can be traced in the analysis of presently available data.


To summarize we have presented a construction of the electron structure 
functions which is the correct approach to lepton
induced processes when the QCD partons are collinear with  the lepton. The
calculation has been done in the leading 
logarihtmic approximation to QCD and leading order in electroweak
interactions. One can definitely  improve this approximation, in 
particular, treating the electroweak sector more precisely.
The $\gamma$-$Z$ interference, contrary to naive expectations,
turns out to be important. Direct calculation of the splitting functions of an electron
into quarks allows for precise control of the momentum scales entering the
evolution. It also shows that the convolution of leptons, electroweak bosons
and quarks should be made at the level of splitting functions rather than 
distribution functions. Unless forced otherwise by the experimental cuts,
the electron splitting functions depend on the external scale $P^2$ and 
influence significantly the parton evolution.

Phenomenological applications of the above analysis require 
very high momentum scales in order to see the weak boson and interference 
contributions. Possible processes where these effects could show up
include heavy flavour, large $p_\perp$ jet and Higgs boson production in lepton
induced processes. At presently available momenta, where the photons dominate, 
the use of the electron structure function allows to  treat correctly  
the parton evolution. It seems also more plausible to use one phenomenological
function of the `resolved' electron instead of parametrizing the 
equivalent bosons' and `resolved' bosons' spectra separately. 

This work has been performed during our visit to Brookhaven National Laboratory and DESY. We would like to thank the 
Theory Groups of BNL and DESY for their hospitality.
One of us (J.S.) thanks the organizers of the Lake Louise Winter Institute
for the invitation to this wonderful meeting.
\eject

\begin{figure}
\centerline{
\mbox{\epsfxsize=0.65\columnwidth \epsfbox{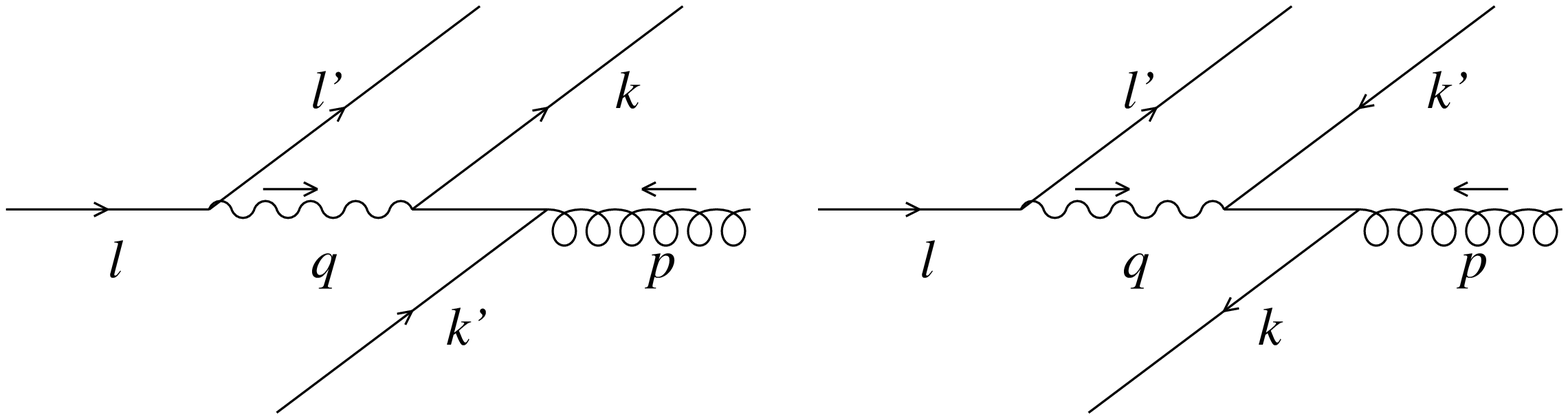}}
}
\caption{Lowest order graphs contributing to the process: $e + G^* \rightarrow \ell' + q + \bar q$.}
\label{F-graf}
\end{figure}

\begin{figure}
\begin{center}
\mbox{\vbox{%
\epsfxsize=7.5cm
\epsfbox{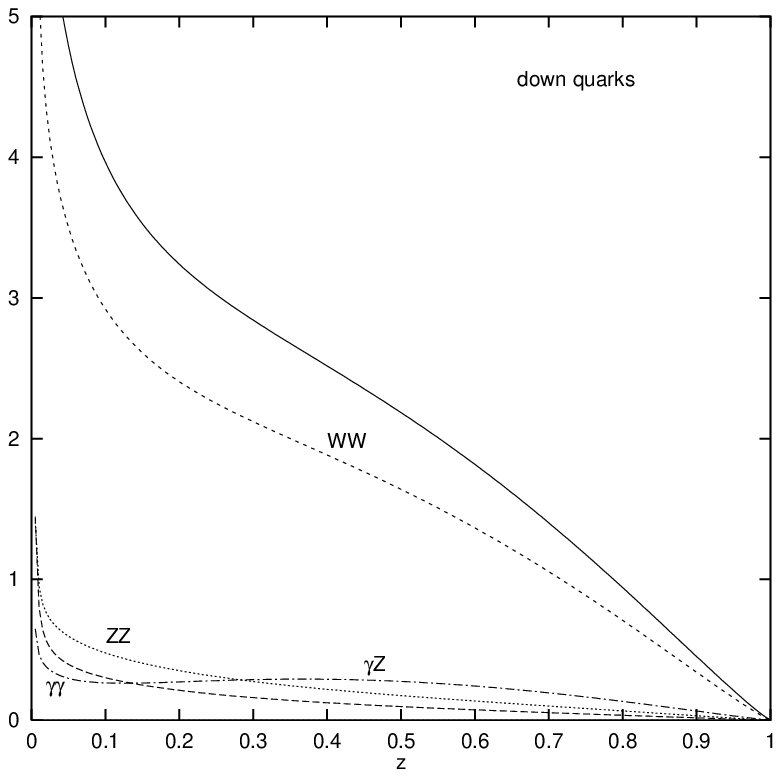}
\epsfxsize=7.5cm
\epsfbox{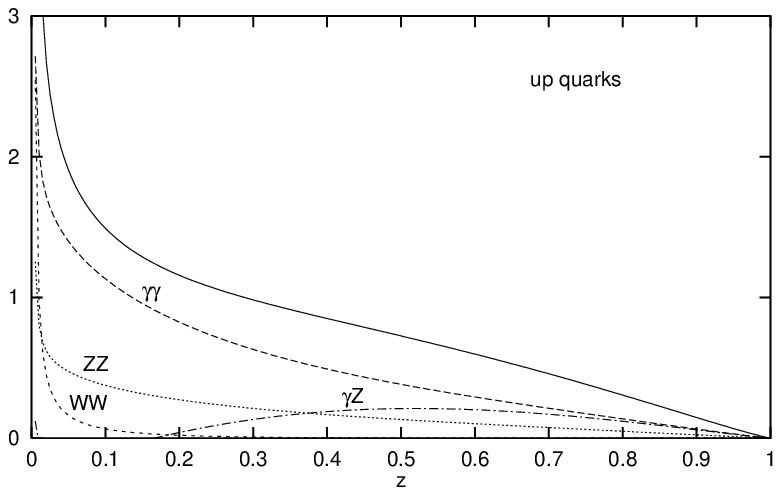}
\epsfxsize=7.5cm
\epsfbox{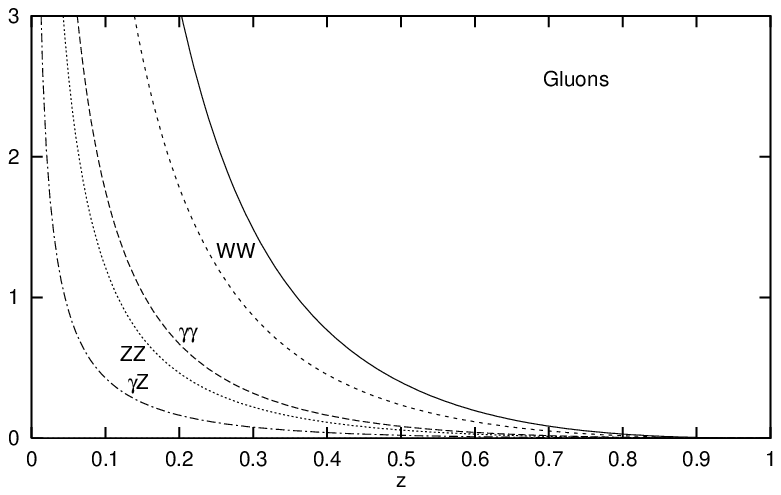}%
}}
\end{center}
\caption{Unpolarized quark and gluon distributions $z f^{\rm as}(z)$ --- solid line.
The other lines show contributions from different electroweak bosons.}
\end{figure}

\begin{figure}[pt]
\begin{center}
\mbox{%
\epsfxsize=8.5cm
\epsfbox{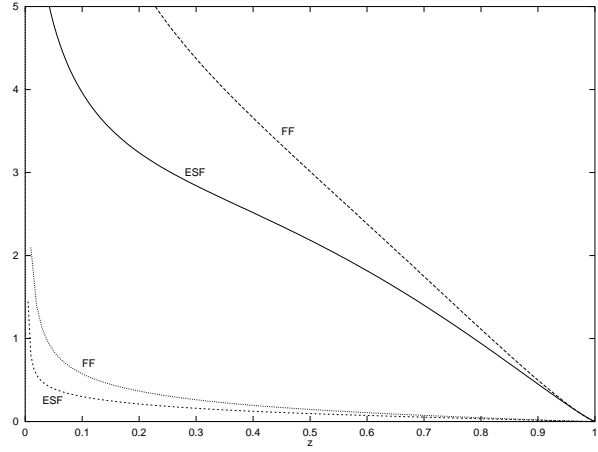}
}
\end{center}
\caption{Comparison
of unpolarized d-quark distributions $z f^{\rm as}(z)$
calculated by ESF and FF methods. The upper two lines result from
contributions from all electroweak bosons while the lower
two from $\gamma\gamma$ only.}
\end{figure}

\end{document}